\documentclass[floatfix,onecolumn,nofootinbib,longbibliography,preprintnumbers,showkeys,10pt]{revtex4}

\usepackage{palatino}

\usepackage{makeidx}
\makeindex

\usepackage{braket} 

\usepackage{graphicx}

\usepackage{subfigure}

\usepackage{amssymb,amsmath,amsthm}

\theoremstyle{remark}

\theoremstyle{remark}

\theoremstyle{remark}

\usepackage{bm} 
\usepackage{graphicx} 
\usepackage{multirow} 
\usepackage{cases} 

\usepackage{booktabs,tabu,fancybox}

\usepackage{verbatim}  

\usepackage{xspace}
\makeatletter
\DeclareRobustCommand\onedot{\futurelet\@let@token\@onedot}
\def\@onedot{\ifx\@let@token.\else.\null\fi\xspace}

\def\eg{\emph{e.g}\onedot} 
\def\ie{\emph{i.e}\onedot} 
 
 \def\vs{\emph{vs}\onedot}

\def\viz{\emph{viz}\onedot}
\makeatother

\usepackage{etoolbox} 
\newtoggle{omitauthor}
\newtoggle{omitwordcount}
\toggletrue{omitwordcount}
\newtoggle{omitdraftnumber}
\toggletrue{omitdraftnumber}
\newtoggle{omitdraftfootnote}
\toggletrue{omitdraftfootnote}

\usepackage{epigraph}

\usepackage[shortlabels,inline]{enumitem}  
\newlist{inlinelist}{enumerate*}{1}  
\setlist[inlinelist]{label=(\roman*)}  
\usepackage{hyperref}

\usepackage{caption}
\usepackage{subcaption}  

\usepackage{chngcntr}

\DeclareMathOperator{\sgn}{sgn}

\newcommand{\numberfield}[1]{\ensuremath{\mathbb{#1}}} 

\newcommand{\vect}[1] {\ensuremath{\mathbf{#1}}} 

\begin{document}

\title{Understanding Quantum Theory: \\ An Operational Reconstructive Approach\footnote{Forthcoming in J. Faye and L. Johansson, \emph{How to Understand Quantum Mechanics: 100 years of Ongoing Interpretations}.}}

\iftoggle{omitauthor}{}
{\author{Philip Goyal}	
    \email{pgoyal@albany.edu}
    \affiliation{University at Albany~(SUNY), NY, USA}
}
\date{\today}
\iftoggle{omitauthor}{}{\homepage[Homepage:~]{https://www.philipgoyal.org}}
\linespread{1.418}

\begin{abstract}
One hundred years after the creation of quantum theory, there is no consensus on the kind of reality that is described by the theory.   In this paper, I attribute the lack of progress to the prevailing interpretative methodology, which invariably takes the quantum formalism as the starting point for philosophical reflection and analysis.  I argue that this methodology is particularly inappropriate for quantum theory.  In particular, it invariably marginalizes much of the theory's content, both that implicit in modelling heuristics and experimental practices, and that encapsulated in the mathematical structures of its formalism.  In addition, the prevailing methodology offers little protection against undue influence by metaphysically-laden language which invariably accompanies the formalism.

Here, I summarize an alternative methodology whose goal is to ensure that an interpretational project take into account all forms of theoretic content.  The methodology harnesses the recent results of the quantum reconstruction program.  These results distil the mathematical content of the quantum formalism into physical principles and assumptions, which are more readily philosophically digestible that the formalism itself, and bracket much of its metaphysically-laden language.   I also outline the main features of quantum reconstruction with a view to highlighting its significance for philosophic interpretation, and describe the value of adopting an operational stance during both reconstruction and interpretation.

As a case study of reconstruction-based interpretation, I describe the reconstruction of the identical particle formalism, and its step-by-step interpretation, highlighting both the key questions that drive the interpretation forwards and the techniques and stances that are employed in each step.  The interpretation yields a novel metaphysical profile for systems of identical particles as \emph{potential parts of a whole}, which can be traced step-by-step to elementary experimental data and the reconstruction's physical postulates and assumptions.  Finally, I describe some of the pitfalls that one faces in any attempt to directly interpret the identical particle formalism.

\end{abstract}


\maketitle

\setlength{\epigraphwidth}{3.5in}
\setlength{\epigraphrule}{0pt}
\epigraph{
	\textit{We are all agreed that your theory is crazy. The question which divides us is whether it is crazy enough to have a chance of being correct. My own feeling is that it is not crazy enough.}
			}%
			{Niels Bohr\footnotemark}   

\section{Introduction}

A physical theory is a bridge between the realm of pure thought---mathematical structures, physical principles, and metaphysical ideals---and immediate sensory experience.  It harnesses systematic, abstract thinking to order sensory experience, and draws upon meticulous observation of natural phenomena to revise and reshape thought.  That such theories exist, that they describe so much of what we observe with such precision, and extend our reach so far beyond the realm of everyday experience, is astonishing, and a testament to the extraordinary powers of the human mind to probe the nature of reality.

A theory is a battle-tested system of thought and practice.  Such a crystalline treasure naturally invites deeper understanding.  A physicist wants to know \emph{why} the theory `works', for example to \emph{extract general features}~(such as metalaws) which can potentially be used to construct new physical theories or reformulate existing ones.  To a philosopher or philosophically-minded physicist, the questions are numerous:~what does the theory `say' about the \emph{nature of reality}?  How does it contrast with the informal metaphysics implicit in our model of the everyday physical world, and with what previous theories appear to say?  Historically, the philosophical interpretation of classical mechanics gave rise to a mechano--geometric--atomistic conception of reality\index{conception of reality!classical}.  This spawned a mechanical world-view\index{mechanical world-view} which, amplified by a string of socially-transformative technologies and other astonishing predictions enabled by classical physics, has profoundly shaped western society over the past four centuries:~our values; our patterns of thought; our beliefs about what exists, counts as real, and is important~\cite{Dijksterhuis1961}.

Since its creation a century ago, quantum theory has not only astonished with its predictive power but has galvanized intense philosophical interest.   Quantum theory radically challenges the ideals, categories, and core metaphysical tenets of the classical conception of reality.   But what ought to replace them?  The sense of mystery---and mystique---of the theory is routinely conveyed in popular accounts of the theory, and is experienced by anyone who has studied the theory at an introductory level.  Here one learns of particles that are supposedly capable of being in two places at once.  Of entangled particles that behave as one even if widely separated.  Of identical particles that crowd together or elbow each other apart.  Of measurements whose outcomes cannot be predicted with certainty and that necessarily \emph{alter} the very physical system they probe.  These ideas are shocking and perplexing to anyone steeped in the presuppositions of classical physics.  Yet one cannot help but take these ideas seriously---because the theory is crazy enough to be true.

A satisfactory philosophical understanding of quantum theory ought to provide a thorough-going \emph{quantum conception of reality}\index{conception of reality!quantum} which provides a \emph{unified} understanding of the above-mentioned---and many other---non-classical features of quantum theory~\cite[\S\,II]{Goyal2022c}.  For example, one would like to have a precise metaphysical profile\index{metaphysics!metaphysical profile} for quantum particles---a description of their time-dependent properties~(if indeed that is a suitable concept)\index{metaphysics!metaphysical profile!quantum properties} as well as a metaphysical profile for systems of identical particles\index{metaphysics!metaphysical profile!identical particles}.  One would like to know the ultimate reason for Bell-inequality violating correlations in entangled particle systems---is it a case of non-local causation, of non-local properties, or something else?  Similarly, one would like to understand why the classical ideal of deterministic causation gives way to probabilistic indeterminism; and why the notion of measurement comes to have such a decisive role in the evolution of quantum systems.

The importance of developing such an understanding can hardly be overstated.  Quantum theory provides an unprecedented opportunity to transform our thinking at the deepest levels, to provide a new conceptual lens on reality which can be broadly propagated.  Given the above-mentioned societal impact of the mechanical world-view, the potential impact of a coherent, detailed, and compelling quantum conception of reality on the future evolution of humanity is incalculable.

Unfortunately, today, a century after the theory's construction, we seem far from any fully-fledged quantum conception of reality, one comparable in detail and comprehensiveness to the classical conception of reality.  Instead, the core quantum formalism has amassed a plethora of widely-divergent interpretations~(the Copenhagen, de Broglie--Bohm, and the Many Worlds interpretations comprising only a few of the available offerings), each narrowly focussed on one or two specific interpretative issues raised by quantum theory~(\eg how to make sense of the measurement postulate).  Philosophical interpretation of the auxiliary components of the quantum formalism is in a similarly unsatisfactory state.  For example, the implications of the identical particle formalism for the metaphysical profile of identical quantum particles is still widely disputed.   A \emph{default interpretation}---one that is compelling and enjoys widespread assent---appears out of reach.  In philosophical circles, such protracted disputes tend to be normalized by the assertion that they are due to underdetermination\index{interpretation!underdetermination}\footnote{See, for example, \cite[\S1.1]{KrauseArenhart2025}.}.  But is such underdetermination fundamental and unavoidable?  That is, must we  accept that philosophical reflection on quantum theory will not yield a compelling default interpretation?\index{interpretation!compelling default}\footnote{For example, the idea that the physical world is populated with \emph{persistent individuals} is our default interpretation.  Metaphysicians can entertain alternative  conceptions, such as \emph{all} of these `objects' are in fact the appearances of a single entity~(monism) or that a `persistent object' is in fact a sequence of momentary objects that arise and pass away in rapid succession. However, such underdetermination appears to be fundamental and unavoidable, at least given everyday sensory experience.  It is for this reason that the default interpretation can hold sway far beyond the scope of academic philosophy---\eg students are not taught that the idea that there exist persistent objects is unscientific and that they should not take it seriously.  In a similar manner, we seek a \emph{default} interpretation of identical quantum particles which can be broadly propagated.}.

What stands in the way of the development of a comprehensive quantum conception of reality?  In my view, the primary challenge one faces in interpreting any battle-tested physical theory is that its content is \emph{polymodal}\index{interpretation!challenge of polymodal content}---\ie distributed across many \emph{different modes of thought}~(mathematical, physical, metaphysical) and \emph{practice}~(modelling heuristics, and experimental practices such as experimental design and analysis of raw data).   This poses a fundamental difficulty:~our minds cannot easily reflect upon knowledge that is manifested in such different forms.  For example, the knowledge contained in modelling heuristics\index{quantum theory!practice of quantum theory!know-how!modelling heuristics} and experimental practices\index{quantum theory!practice of quantum theory!experimental practices} is largely \emph{know-how}\index{quantum theory!practice of quantum theory!know-how}, which is learned by doing; it is rarely made explicit, even in textbooks or research literature; and is typically rather \emph{ad hoc}.  Similarly, a theory's mathematical structures encapsulate a large number of mathematically-driven choices, whose \emph{physical} meaning is often opaque.

In the case of, say, classical mechanics, the polymodal content is somewhat manageable since  \begin{inlinelist} \item the theory was built on metaphysical ground that was carefully staked out and argued for; \item  its mathematics substantially flows from readily intelligible physical principles, with a minimum of appeal to mathematical aesthetics; and \item the theory concerns objects that can be directly manipulated or observed~\cite[\S\,II]{Goyal2022c}. \end{inlinelist}    However, none of these conditions hold in the case of quantum theory~\cite[\S\,III.A]{Goyal2022c}.   For example, quantum theory was certainly \emph{not} built up on a carefully staked out metaphysical foundation.  Similarly, the development of its mathematical structures involved substantial appeal to mathematical guesswork and aesthetic considerations.  It is in this context that the prevailing interpretative methodology\index{interpretation!methodology!prevailing}, filtered by various cognitive biases\index{interpretation!cognitive biases}~\cite[\S~II]{Goyal2025a}, is ill-suited for the task at hand.

The first limitation of the prevailing interpretative methodology is that it \textit{focusses on the abstract quantum formalism}, marginalizing or neglecting other forms of content~(especially that which is implicit in modelling and experimental practices)\index{interpretation!methodology!marginalization of implicit content}.   This represents a \emph{bias towards the explicit and general} over the implicit and heuristic\index{interpretation!cognitive biases!explicit over implicit}\index{interpretation!cognitive biases!general over heuristic}. Second, it is standard practice to \emph{neglect most of the content of the abstract quantum formalism} itself.  This is a case of  \emph{availability bias}\index{interpretation!cognitive biases!availability bias}, which manifests as a focus on \begin{inlinelist} \item the natural-language component of the formalism\index{interpretation!methodology!focus on natural-language component}, or on that portion of the formalism that can be rendered in intelligible form~(for example, via reformulation of the theory, as in Bohm's Hamilton-Jacobi reformulation of the Schr\"odinger equation); and \item the neglect of opaque mathematical structures. \end{inlinelist}  Hence, the bulk of the mathematical structure of the formalism is either left uninterpreted, or its interpretation is subordinated to the natural-language component.

Hence, the viability of the prevailing interpretational method rests on the implicit assumption that an interpretation so generated will not be undermined by whatever may be the content of the theory's modelling and experimental practices or of its mathematical structures.  However, as demonstrated by the overthrow of the aether interpretation of Maxwell's equations by Einstein's derivation of the Lorentz transformations~\cite[\S\,III.A]{Goyal2025a}, this is an \emph{unsafe} assumption~\cite[\S\,IV.C]{Goyal2025a}.

In recent work~\cite{Goyal2022c, Goyal2025a}, I propose a new reconstruction-based interpretative methodology\index{interpretation!methodology!reconstruction-based interpretation} that strives to better meet the challenge of polymodal theoretic content.  It based on three core ideas:
\begin{enumerate}
\item  Interpret \emph{all} parts of the practice of quantum theory\index{quantum theory!practice of quantum theory}---\ie its entire theoretic content, whether crisp formalism, modelling heuristics, or experimental practices.  To assist, construct an \emph{interpretation-free zone}\index{interpretation!interpretation-free zone} around the theory.  Within that zone, adopt a \emph{descriptive} rather than an \emph{interpretative} stance\index{interpretation!interpretation-free zone!descriptive stance}, which helps to counter the sorts of cognitive biases mentioned above.  Once the descriptive process has been completed, make explicit judgements as to what facts are and are not interpretationally relevant.  Interpretation can then take place on the basis of the facts deemed relevant.  Assessment by others can then take into account not just the proffered interpretation but also why certain facts were deemed relevant but others were not.

\item Reflect on \emph{reconstructions} of the parts of the quantum formalism of interest, rather than on the formalism \emph{per se}.  This has many benefits, including surfacing more of the theory's physical content for philosophical reflection, bracketing metaphysical connotations of the language typically used to talk about the formalism, and provoking new interpretational possibilities.

\item Use \emph{operational analysis} to properly ground all parts of the formalism in experimental practice.  The aim here is to surface key assumptions of interpretative import that are implicit in experimental practice, and to identify concepts that are not well grounded in the bare experimental data but are typically regarded as integral to the formalism.
\end{enumerate}

The methodology takes advantage of the extraordinary recent successes of the \emph{quantum reconstruction program}\index{quantum reconstruction!quantum reconstruction program}.    For most of quantum theory's  life, reconstructions---systematic, principled derivations---of its formalism have been unavailable, and indeed were thought to be out of reach.  However, over the past twenty-five years, most components of the quantum formalism have been reconstructed---\begin{inlinelist} \item the finite-dimensional von Neumann--Dirac axioms, from a wide variety of angles~(\eg~\cite{Hardy01a,Hardy01b, Goyal-QT2c,GKS-PRA, DakicBrukner2010, Dariano-operational-axioms,Chiribella2011, Goyal2014, Hohn2017, MullerMasanes2016, MasanesMullerAugusiakPerez-Garcial2013,SelbyScandoloCoecke2018}); \item the quantum wave equations~(e.g.~\cite{DarianoPerinotti2014,Goyal-correspondence-rules-2010}); \item the identical particle formalism~\cite{Goyal2015,Goyal2019a,SanchezDakic2024}; and \item the classical-quantum correspondence rules~\cite{Goyal-correspondence-rules-2010}. \end{inlinelist}   These reconstructive results offer a marvellous opportunity to approach interpretation with renewed vigour.   

Two interpretational case studies that employ the above methodology now exist:
\begin{enumerate} 
\item A study of the nature of identical particles~\cite{Goyal2019a,Goyal2023a,Goyal2025b}, based on a reconstruction in~\cite{Goyal2015}\index{quantum reconstruction!identical particles}.
\item  A study of the properties of quantum particles~\cite{Goyal2025c}, based on a reconstruction of the finite-dimensional von Neumann axioms~\cite{Goyal2014} via a reconstruction of Feynman's formalism~\cite{GKS-PRA}\index{quantum reconstruction!reconstruction of Feynman's formalism}. 
\end{enumerate}  
These case studies yield novel interpretations of identical particles and quantum properties.  In addition, the former provides detailed insight in the pitfalls of directly interpreting a formalism, and the consequences of not taking due account of modelling heuristics and experimental practices.

The remainder of this paper is organized as follows.  First, despite its many successes, the reconstructive methodology\index{reconstructive methodology} remains unfamiliar and minimally utilized in philosophical circles.  Accordingly, I shall begin with some remarks and examples on how the reconstructive methodology has been used historically in both physics and mathematics\index{reconstructive methodology!historical use in physics}\index{reconstructive methodology!historical use in mathematics}~(Sec.~\ref{sec:reconstruction-examples}); and then discuss how, when, and why a reconstruction can have ramifications for the philosophical understanding of a theory~(Sec.~\ref{sec:reconstruction-philosophical-impact})\index{quantum reconstruction!philosophical ramifications}.   

Second, the operational stance\index{operationalism!operational stance} is key to most reconstructions, and also to the reconstruction-based methodology outlined above.  Yet, the operational stance is often regarded as an anathema in philosophical circles, at least in the context of interpreting quantum theory.  Accordingly, I shall discuss the role of this stance in the creation and interpretation of physical theories\index{operationalism!operational stance!role in theory creation}\index{operationalism!operational stance!role in interpretation}, and shall make some remarks in the hope of dispelling some potential misunderstandings\index{operationalism!operational stance!misconceptions} about the operational stance~(Sec.~\ref{sec:operationalism}). 

Third, although reconstructions have been available for close to twenty-five years, very little effort has been directed towards their interpretation.  This suggests a continuing, pervasive uncertainty about the interpretative potential of reconstructions, and also the absence of a clear methodology that indicates how to go about mining a reconstruction for interpretive insight.  In my view, both of these obstacles are most effectively addressed through examination of case studies.  Accordingly, in Sec.~\ref{sec:identical-particles}, I shall examine the above-mentioned reconstruction-based interpretation of the quantum identical particle formalism, with a view to pointing out the key questions that drive the investigation forwards, and the techniques and stances~(in particular the operational stance) that yield answers to these questions.  I shall also indicate how, in this case, this interpretational process helps to reveal many pitfalls in the prevailing interpretative methodology.

\section{Reconstruction of Quantum Theory}

\subsection{Importance of Reconstruction in the History of Physics and Mathematics}
\label{sec:reconstruction-examples}

A physical theory is expressed in terms of mathematical structures and mathematical symbolism.  When the theory is being constructed, these mathematical structures are never fully determined by metaphysical posits and physical ideas or principles. Rather, as these ideas and principles are being clothed in mathematical forms, \emph{choices} are made on the basis of \emph{mathematical} considerations, such as symbolic simplicity, symmetry, and the availability of relevant mathematical structures.  Classical mechanics furnishes many simple examples of this phenomenon\index{reconstructive methodology!historical use in physics!classical mechanics}.  For instance, the idea that a moving body possess a \emph{quantity of motion} was translated by Descartes\footnote{Descartes asserts: ``there is a fixed and determined quantity of [motion] ...always the same in the universe as a whole even though there may at times be more or less motion in certain of its individual parts'', and that ``when one part of matter moves twice as fast as another twice as large, there is as much motion in the smaller as in the larger''.~\cite[II 36]{Descartes1644}.    Taking size as a proxy for mass, this translates into the conserved quantity~$\sum_i m_i u_i$, where~$m_i$ is a measure of the `size'~(\viz \emph{volume}) of body~$i$.} into a \emph{mathematical quantity}, the scalar~$mv$.  Presumably this choice was recommended by its symbolic simplicity.  Descartes presents it as if it were self-evident, and offers no further justification for his mathematical choice, even though an infinity of others---$f(m,v)$, with~$f$ monotonic increasing in~$m, v$---would express the same general idea.  Similar mathematical choices are widespread in quantum theory.  In Schr\"odinger's formulation of quantum theory, the use of complex numbers, the representation of $n$-particle states in $3n$-dimensional configuration space, and the posited form~$\left|\psi(\vect{r},t) \right|^2$ of the Born rule, are all \emph{choices} coloured by \emph{mathematical} considerations.  Again, many other choices are conceivable.

During the \emph{developmental} phase in the life of a theory, such mathematical choices are justified \emph{post-hoc} by the fruits of the theory to which they give rise.  However, in the subsequent \emph{reflective phase}, understanding of the well-proven theory can be deepened by deriving its mathematical forms in a systematic manner from simpler mathematical assertions that one can more fully justify on purely physical grounds\footnote{For further discussion on this point, and for a discussion of other questions that are typically raised in the reflective phase, see~\cite{Goyal2022c}.}.  Such derivations thereby transmute opaque \emph{mathematical} choices into \emph{physical} content, which can then be more readily understood.  Classical mechanics furnishes many simple examples of such derivations:
\begin{enumerate}[(i)] 
\item d'Alembert's derivation of Newton's parallelogram of forces\footnote{See~\cite[\S\,IV.A.2]{Goyal2023a} for elaboration, \cite[Ch.~1]{AczelDhombres1989} for mathematical details, and~\cite{Lange2011} for historical context.}; 
\item Derivations of momentum and kinetic energy in non-relativistic and relativistic physics\footnote{See~\cite{Goyal2020} for derivations, historical context, and extensive citations.}; and
\item Derivation of Newton's second law\footnote{See~\cite{Darrigol2019} for historical context, and~\cite[\S2.3]{Goyal2020} for a contemporary derivation.}.
\end{enumerate}
As described in~\cite[\S\,III.A]{Goyal2025a}, the enhancement in understanding provided in these instances can be attributed to \emph{stratification}---tracing back of the given formalism to assumptions and principles possessing a higher degree of simplicity, which are more readily intelligible or justifiable on the basis of metaphysical or physical axioms, or on the basis of physical principles~(either existing or novel).   For example, d'Alembert's derivation traces Newton's parallelogram of forces~(which Newton had justified by appeal to motion and impulses) to fundamental physical symmetries~(isotropy of space) and combinatorial symmetries~(associativity and commutativity of combination of forces).  Indeed, the resulting derivation is so general that it applies to abstract geometric vectors.  Similarly, the derivation of the quantities of motion in~\cite{Goyal2020} relies on compositional symmetry~(associativity), spatial isotropy, and continuity, together with the principle of relativity and a generic principle of conservation.  The derivation thereby traces the particular mathematical forms posited by Buridan, Descartes, and others to fundamental geometrical, combinatorial, and physical symmetries.  Moreover, the derivation generalizes trivially to special relativistic dynamics, thereby placing classical and relativistic quantities of motion on a common footing.

Reconstruction is also employed in mathematics to illuminate useful but puzzling constructs\index{reconstructive methodology!historical use in mathematics!complex numbers}.  For example, complex numbers were introduced in 1545 by Cardano as a book-keeping device in the solution of the cubic equation.  Once the utility and deep unifying power of complex numbers was established over the subsequent 250 years, the question arose as to their logical foundations. Hamilton's 1831 formulation of complex numbers as real number pairs subject to basic compositional symmetries~(associativity, distributivity, commutativity) and basic algebraic requirements~(multiplicative norm, possibility of division), all of which were abstracted from real number arithmetic, represented the first critical step towards reconstruction~\cite{Hamilton1831}.  This paved the way for the fundamental theorems of Frobenius~\cite{Frobenius1878} and Hurwitz~\cite{Hurwitz1898}, which \emph{reconstruct} complex numbers, in the sense of showing precisely how complex number arithmetic arises from fundamental compositional symmetries and basic algebraic requirements.  Moreover, their work lead to the discovery and precise delineation of a broader landscape of generalizations of real number arithmetic---quaternions~(subsequently leading to Clifford algebra and geometric algebra) and octonions---in which complex numbers are situated.

\subsection{Impact of Reconstruction on Philosophical Understanding}
\label{sec:reconstruction-philosophical-impact}

The impact of reconstruction\index{reconstructive methodology!impact} is often purely \emph{internal}\index{reconstructive methodology!impact!internal}, confined to the field it which it is applied~(\eg physics or mathematics).  However, it sometimes also has \emph{external} impact\index{reconstructive methodology!impact!external}, extending to the layer of metaphysical posits, categories, and overarching desiderata which shape enquiry in the field as a whole.  The above-mentioned reconstructive examples concerning classical mechanics and complex numbers are an example of purely internal impact.  However, Einstein's reconstruction of the Lorentz transformations is a case of \emph{both}---it lead to the reconsideration of the absolute Newtonian distinctions between space and time and between matter and interactions\index{reconstructive methodology!historical use in physics!Lorentz transformations}.  These reconsiderations not only opened up the space for new developments in physics~(\eg general relativity, with its curved spacetime manifold; and matter--energy transformations), but resculpted the basic categories into which reality is divided~\cite[\S\,III.A]{Goyal2025a}.

Is it possible to understand why Einstein's reconstruction had such broad impact, whereas those in, say, classical mechanics had purely internal impact?  Perhaps the singular point of difference is that the target of Einstein's reconstruction was not a mathematical structure that had been directly guessed~(such as the forms of the classical quantities of motion), but rather a mathematical structure that arose through a complex fermentation spanning more than sixty years---a symmetry group emanating from Maxwell's equations, which in turn are a mathematical integration of at least \emph{four} distinct physical principles or laws.

The fact that the mathematical structures employed in the quantum formalism arose in a convoluted manner through a mathematical integration of various physical ideas certainly raises the expectation that a reconstruction of these structures will lead to new philosophical interpretative insights or, indeed, to new interpretations.  But whether this is likely---and thus is deserving of serious effort---cannot be decided \emph{a priori}.  For one thing, it will depend on the particular reconstruction in question.  In my view, there is no alternative to simply \emph{trying}---to philosophically reflect on well-chosen reconstructions and to see what arises.  As discussed below, the early signs are promising.

\section{The Operational Stance}\index{operationalism!operational stance}
\label{sec:operationalism}

A physical theory employs many concepts that can be viewed as refinements of informal metaphysical ideas that are integral to our everyday conception of the physical world.  For example, Euclidean geometry refines~(precisifies and extrapolates)  the informal notions of location and straightness; and Newtonian physics refines the informal notion of \begin{inlinelist} \item `the same time', \item the `object + property' schema~(substance and attributes), and \item `the same object'~(transtemporal identity).\end{inlinelist}  In each such instance, one can conceive of other refinements of a given common notion.   For example,  `straightness' could be extrapolated as a global notion~(infinitely-long straight line, as in Euclidean geometry) or a local one~(locally-straight curve, as in Riemannian geometry).  However, once a refined notion has achieved precise expression, and has shown its utility within an empirically-validated theory, it seems to acquire a nearly immutable status. 

The adoption of a refined concept is thus a double-edged sword.  On the one hand, it offers clarity and uniformity, which can be used to build elegant, formalized systems of thought~(such as Euclidean geometry or Newtonian mechanics).  On the other hand, one runs the risk of extrapolating everyday notions farther than is empirically warranted.  Accordingly, the apparent immutability or indispensability of well-established formalized concepts can interfere with the creation of a new theory, or with the proper interpretation of an existing theory.  For this reason, we need methods that assist us in identifying instances of overreach.   An operational stance towards a theory is perhaps the most powerful method available.  

In taking an operational stance, one adopts the view that one has more reason to grant credence to the bare data of sensory experience than to abstract concepts~(whether they be everyday concepts or refined versions thereof).  Of course, `bare data' is an idealization---in reality, all observation statements\index{operationalism!observation statements} are colored by experience, language, and culture.  However, it is possible to identify some general criteria for \emph{differentiating} between observation statements---\ie for regarding certain observation statements as closer to `bare data' than others\index{operationalism!observation statements!differentiation}.   For example, insofar as we regard ourselves as embodied beings of finite extent, direct sensory experience is necessarily spatially and temporally \emph{local}.  Hence, we have more reason to grant credence to observation statements regarding such experiences than to those statements that extrapolate beyond them.  

For instance, observation statements regarding what patterns of colour I experience in my visual field at this moment are more reliable than statements asserting what patterns I see in the distance~(which assertion refers to spatially non-local events) or what \emph{persistent objects} I see in my immediate vicinity~(which assertion integrates momentary experience across time via an object-model of what underlies such momentary experience).   A historically pivotal example of such a distinction occurs in Poincar\'e's analysis of the experimental determination of the simultaneity of events~\cite{Poincare1898}, which establishes that a judgement of the simultaneity of distant events\index{operationalism!Poincar\'e's simultaneity analysis} rests upon assumptions that cannot be directly and separately tested~\cite[\S\,II.B.2.3]{Goyal2025b}.  The upshot of this analysis is that we ought to take observation statements about the simultaneity of local events~(\ie infinitesimally-close events) as the baseline~(presumably valid across an inconceivably broad range of theories), and to regard those about distant events as strongly theory-dependent.

The operational stance also cultivates sensitivity to the risk of error inherent to the \emph{extrapolation} of concepts from the sphere of physical phenomena in which their utility is established to a new sphere that manifestly differs from the former\index{operationalism!risk of concept extrapolation}.  For example, the electron orbits posited by Bohr in his model of the atom represented an extrapolation of the Newtonian abstraction of a particle trajectory from the macroscopic realm to the microscopic.  Although Bohr's model was effective, its predictive power was limited, as were those of outcomes to attempt to directly generalize it~(\eg Sommerfeld's elliptical orbits).  Prompted by Einstein's operational stance, Heisenberg recognized that the very posit that an electron travels in an orbit is a questionable extrapolation as we have no \emph{direct} sensory evidence that such a posit is valid.  Hence, Heisenberg instead focussed on the what is actually observed, namely spectral line frequencies, and  discovered that it was possible to develop an astonishingly capable calculus for predicting this data~(and a great deal else besides) without risking use of this questionable posit\index{operationalism!risk of concept extrapolation!orbits in Bohr's atomic model}.

\subsection{Operational Stance in Quantum Reconstruction}\index{operationalism!quantum reconstruction}

An operational stance is default in the quantum reconstruction program.  From this stance, the abstract quantum formalism~(\viz the Dirac--von Neumann axioms) is viewed minimally as a means to connect the outcomes of successive measurements performed on an abstract quantum system.   More precisely, an \emph{operational framework}\index{quantum reconstruction!operational framework} is formulated, wherein one presumes that one can partition physical reality into \emph{system} and \emph{non-system}, and can sub-partition the latter into \emph{environment}, \emph{observer}, \emph{agent}, and \emph{rest of the universe}~\cite[\S\,II.B.2]{Goyal2025b}.   One makes a number of additional assumptions that idealize those implicit in experimental practice, such as the agent being free to act without being influenced by the system under study.  An experiment is then taken to be performed on a system in an environment with which it interacts, and which undergoes \emph{measurements} that yield \emph{outcomes}.

As in Heisenberg's approach to formulating quantum mechanics, the intent in such an operational stance is to \emph{bracket} physical concepts that may not be applicable~(at least in unmodified form) to the quantum realm, and to instead focus attention as far as possible on reconstructing quantum theory on the basis of postulates that only refer to what an idealized experimenter controls~(design of experiments and the settings of macroscopic apparatus) and experiences~(detections in macroscopic apparatus, idealized as measurement outcomes)\index{operationalism!bracketing physical concepts}.

In the reconstruction of auxiliary parts of the quantum formalism, such as the classical-quantum correspondence rules and the identical particle formalism, one must additionally identify and analyse the experimental procedures that instantiate the operational framework.   For example, in reconstructing the position-momentum commutation relation,~$[\hat{x}, \hat{p}] = i\hbar$, one must consider the procedures that are to be used in measuring the quantities represented by these operators~\cite{Goyal-correspondence-rules-2010}.

\subsection{Operational Stance in Interpretation of Quantum Reconstructions}\index{operationalism!interpretation of quantum reconstructions}

The operational stance in quantum reconstruction brackets many of the metaphysical notions that are commonly associated with the theory.  Consequently, it provides a radically different vantage-point for interpretation~\cite[\S\,VI.A]{Goyal2022c}.

For example, the quantum formalism is ordinarily stated in terms of the \emph{state} concept, which is usually interpreted in the same manner as in classical theories, namely as referring to an \emph{observer-independent physical state} of a physical object.  However, in operational reconstructions, a mathematical state is introduced solely as a means to connect outcomes of measurements, and to enable the prediction of outcome probabilities of measurements performed on the system.  From this point of view, the state is best viewed as descriptive of a reality \emph{as probed by} an ideal agent--observer\index{operationalism!interpretation of quantum reconstructions!quantum state}.

The operational stance also brackets some of the \emph{categories} into which reality is usually carved.  This bracketing allows us to discover the categories upon which the quantum formalism actually depends.  For example, operational reconstructions typically make no use of postulates that refer to space.  That is, the abstract quantum formalism can be derived without any reference to the existence of space, and certainly to its dimension, topology, or metric.  However, \emph{time} plays a necessary role since measurements and interactions must be ordered in time.  The interpretational implication is that, in light of quantum theory, \emph{time} is fundamental, but \emph{space} is secondary.  This fact, together with the existence of Bell-violating correlations due to entanglement, point to space as a secondary, perhaps emergent structure\index{operationalism!interpretation of quantum reconstructions!primary of time over space}.

Interpretation of reconstructions of the abstract quantum formalism often requires additional operational analysis.  For example, in the derivation of the tensor product rule~(which determines the state of a composite system given the pure states of its subsystems), one usually interprets the component subsystems as \emph{persistent individuals}.  However, once these components are put in an \emph{entangled} state, is this interpretation still warranted?  Here, an operational analysis is needed in order to determine what are the experimental procedures which enable component reidentification, on which basis one can assess whether one can retain the individuality interpretation of the components if the composite is in an entangled state\index{operationalism!interpretation of quantum reconstructions!reidentifiability}.

The auxiliary parts of the quantum formalism, such as the classical-correspondence rules and the identical particle formalism, typically involve more extensive contact with the formalism and conceptual framework of classical theories.  For example, the commutation relations explicitly refer to properties such as position and momentum\index{operationalism!interpretation of quantum reconstructions!position \vs momentum}.  Similarly, the identical particle formalism refers to identical particles, \ie particles with identical time-independent properties~(such as mass and charge), thereby differentiating between the time-independent and time-dependent properties of particles.   In such cases, an operational analysis is indicated in order to guide interpretation of the reconstructions of these parts of the formalism.      For example, in formulating a  metaphysical model of quantum properties on the basis of a reconstruction, an operational analysis allows time-dependent properties to be differentiated by their degree of theory-dependence.  For instance, one finds that measurement of position---an instantaneous categorical property---can be regarded as a  baseline, common to classical physics and quantum theory; whereas procedures to measure momentum are strongly theory-dependent~\cite{Goyal2025c}.

\subsection{Misconceptions about Operationalism}\index{operationalism!misconceptions}

As mentioned in the Introduction, there is a general tendency to marginalize the theoretic content of experimental practices.  This has lead to the widespread interpretational attitude that the theories of classical physics say nothing about measurement, and hence speak directly about what nature is like.  Accordingly, the quantum measurement postulate is viewed as an impediment to interpreting quantum theory in the same manner, and is often treated as if it were a defect in the formalism.  This viewpoint is reflected in most of the popular interpretations, which each attempt to demote the measurement postulate~(and its associated probability concept) in some manner or other whilst retaining and elevating the unitary part of the formalism\index{operationalism!misconceptions!measurement postulate as impediment}.

However, any physical theory capable of experimental testing contains---either implicitly or explicitly---a model of the measurement process.  Here \emph{measurement} should be understood as an idealized, abstract process wherein an \emph{interaction} with a \emph{system}  yields an observable \emph{outcome}, and thereby serves as an interface between a physical system of interest and the senses of an observer.   In the theories of classical physics, ideal measurements are presumed to be \emph{passive}~(\ie yielding information about the system under observation without changing the system in any way) and \emph{exhaustive}~(\ie yielding \emph{all} information about the state of the system).  Accordingly, the \emph{formalism} of the theory has no need to explicitly mention this fact.  Rather, this model of the measurement process is incorporated into \emph{experimental procedures}.    In contrast, any theory, such as quantum theory, in which ideal measurements are deemed to be \emph{active} or \emph{non-exhaustive}, must contain a `measurement postulate' in its \emph{formalism}.  Hence, no physical theory---whether classical or quantum---provides a measurement-free window on reality.  

A reason commonly given for demoting the quantum measurement postulate is that the quantum formalism does not itself specify what type of interactions constitute a measurement\index{operationalism!misconceptions!self-description of measurement}.  However, it is not necessary that the formalism of a physical theory self-describes its own measurement procedures.  No extant theory, such as classical mechanics, meets this requirement.   Indeed, meeting such a requirement may be a logical impossibility.  In any case, it is sufficient that the formalism is accompanied by experimental procedures that provide solid guidance in identifying and implementing processes that implement measurements.  In particular, while it is true that the quantum formalism provides no guidance on identifying what kinds of processes constitute measurements, ample guidance is provided by the body of experimental practices associated with the theory.  For example, in a quantum optics experiment, it is well-established that certain elements of the apparatus~(such as beam splitters and mirrors) can be modelled as unitary transformations on the quantum system, while others~(such as detectors) must be modelled as measurements.  Such know-how is perfectly sufficient for quantum optics as an experimental and theoretical field.   This does not preclude the development of physical models of the measurement process~(such as Penrose's gravitationally-induced collapse or GRW's collapse model) which can provide more guidance.  But such models---and certainly not perfect ones---are not a prerequisite for the quantum measurement postulate to be taken seriously on its own terms.

\section{Identical Quantum Particles: Reconstruction and Interpretation}\index{quantum reconstruction!identical particles}\index{identical particles}
\label{sec:identical-particles}
The nature of identical quantum particles has been in question since the formulation of Bose-Einstein and Fermi-Dirac statistics, and Heisenberg and Dirac's incorporation of these statistics into the nascent quantum formalism via the so-called symmetrization postulate.  On the basis of these statistics and formalism, two interpretations were put forward by the founders of quantum theory.  Dirac asserted that identical particles are \emph{indistinguishable}, while Schr\"odinger proposed that identical particles are not individuals in the sense that they are deprived of absolute transtemporal identity.  Although the topic of identical particles has been of intense interpretational interest for several decades, it is beset by metaphysical underdetermination.  The options laid out by Dirac and Schr\"odinger remain the two main metaphysical packages on the table~\cite{French2000}.

The methodology invariably employed to establish the nature of identical particles is to directly interpret a quantum formalism for handling identical particles---usually the Dirac symmetrization procedure;  less commonly the Fock space representation of this procedure.  As indicated in the Introduction, such a methodology is inherently problematic.  In this case, the problems manifest in several ways, some of which will be described below.   

In the reconstructive approach to interpreting the identical particle formalism, one does not philosophically reflect on the formalism directly, but rather on a \emph{reconstruction} of the identical particle formalism---\ie on the postulates and framework used in the reconstruction.  In this case, the reconstruction in question is carried out in the context of Feynman's formulation of quantum theory, and accordingly yields the Feynman form of the symmetrization postulate~\cite{Goyal2015}.  Its interpretation has been built up in three distinct steps over the course of several years, and is described in three corresponding publications:
\begin{enumerate}
\item\emph{Persistence \& Nonpersistence as Complementary Models of Identical Quantum Particles}~\cite{Goyal2019a}.
\item\emph{Persistence and Reidentification in Systems of Identical Particles}~\cite{Goyal2023a}.
\item\emph{Identical Quantum Particles as Potential Parts}~\cite{Goyal2025b}.
\end{enumerate}
The aim of the discussion below is two-fold.  First, to indicate what techniques or stances~(such as the operational stance, and attending to modelling heuristics) are used to make headway at each step.  Second, to articulate the key open questions at each step which propel the investigation forward.

\subsection{Reconstruction of the Feynman Symmetrization Procedure}\index{quantum reconstruction!identical particles!Feynman symmetrization procedure}\index{identical particles!reconstruction}
\label{sec:reconstruction-SP}

As posited by Dirac~\cite{Dirac1926} and Heisenberg~\cite{Heisenberg1926}, the symmetrization postulate is an essentially \emph{ad hoc} mathematical device for constructing viable models of systems of identical particles\index{quantum reconstruction!identical particles!\emph{ad hoc} symmetrization postulate}.  Even after the empirical success of the symmetrization postulate, Dirac asserted that `other more complicated kinds of symmetry are possible mathematically, but do not apply to any known particles'.  This remark reflects the fact that, although the path to the symmetrization postulate was \emph{guided} by physical ideas~(such as `indistinguishability'), giving it precise mathematical form required that \emph{mathematical} choices be made, choices which could not be justified \emph{a priori}, but only \emph{a posteriori} by appeal to empirical adequacy.  In a similar vein, Pauli in his 1946 Nobel lecture~\cite{Pauli1946} expressed the view that his exclusion principle~(even after incorporation into the symmetrization procedure) called for a deeper explanation:
\begin{quote}
Already in my original paper I stressed the circumstance that I was unable to give a logical reason for the exclusion principle or to deduce it from more general assumptions. I had always the feeling and I still have it today, that this is a deficiency. Of course in the beginning I hoped that the new quantum mechanics, with the help of which it was possible to deduce so many half-empirical formal rules in use at that time, will also rigorously deduce the exclusion principle. Instead of it there was for electrons still an exclusion: not of particular states any longer, but of whole classes of states, namely the exclusion of all classes different from the antisymmetrical one.'~\cite[p.~32]{Pauli1946}.
\end{quote}
With quantum theory in its reflective stage of life, the symmetrization postulate is ripe for elucidation via reconstruction.

\subsubsection{Adopting an Operational Stance}\index{quantum reconstruction!identical particles!operational analysis}\index{identical particles!operational analysis}
The first step in an operational reconstruction is to zero in on the \emph{primary}~(`bare') experimental facts, and to differentiate these from conceptual embellishments thereof which are implicit in an experimentalist's observation statements.  In the case of systems of identical particles in, say, a bubble chamber, the primary data consists images consisting of bubbles arranged in various curved segments.  A key step in the usual experimental analysis of such an image is to associate each curve-segment with a `particle', minimally understood as a persistent individual that interacts with a fluid to generate an observable outcome~(a bubble).  Thus, a particle physicist will speak of `particle tracks', rather than bubbles or elementary detections, even though the former are higher-order notions.  

Now, the particle association is clearly essential to particle physics since it \eg enables information about the time-independent properties of the particle to be calculated from the geometry of the associated curve-segment.  However, upon adopting an operational stance, one notes that, unlike the case with, say, observation of an aeroplane in the sky~(where one observes not just the aeroplane's contrail but the aeroplane itself), one never \emph{directly observes} a particle---one observes a \emph{sequence of bubbles}.    In addition, what one treats as a particle track is not continuous, but composed of a \emph{discrete} sequence of visibly-separated bubbles.  This contrasts with a classical-conceived contrail, or the trajectory of a classical mechanical particle, whose exact, continuous observation is not precluded by classical physics.  For these reasons, one recognizes that the customary association of particle with curve-segment is on shakier ground than, say, the association of an aeroplane with a contrail.

\subsubsection{Exploiting a Gap Identified by Operational Analysis}\index{quantum reconstruction!identical particles!bracketing individual persistence}\index{identical particles!bracketing individual persistence}
\label{sec:exploiting-a-gap}
The reconstruction in~\cite{Goyal2015} seizes upon the gap between the bare experimental facts and the way in which they are usually parsed.  The essential point is that, given two detections at locations~$\ell_1$ and~$\ell_2$ at time~$t_1$, and two detections at locations~$m_1$ and~$m_2$ at time~$t_2$, we have reason to be wary of assertions about what \emph{happened} in between given that we do directly observe what has happened.  

In particular, in~\cite[\S1]{Goyal2015}, concern is raised about the validity of statements concerning the occurrence of transitions unseen by us.  The rationale given refers to the widespread notion that identical particles are indistinguishable:~``Even the conclusion that `the particle that generated outcome $\ell_1$ was also responsible for outcome $m_1$ or $m_2$, but we do not know which' is invalid since we cannot verify this statement on account of particle indistinguishability.''  We will revisit this rationale in due course, find it wanting, and replace it with a more sturdy one~(\S\ref{sec:interpretation-step-1}).  However, for the present, the essential point is that one recognizes that an assertion about unseen transitions is \emph{questionable} since it relies upon classical intuitions about objects that may not be applicable in the present context~(these will be investigated in~\S\ref{sec:interpretation-step-1}).    Accordingly, the reconstruction adopts the strategy of \emph{bracketing} this assertion, and then seeks to derive a Feynman transition amplitude from the initial to final detections without it.  This move is analogous to Heisenberg's eschewance of Bohr's electronic orbits in his derivation of the matrix mechanics:~although Heisenberg did not give a positive reason for thinking that the notion of Bohrian orbits was \emph{invalid}, he did have reason to be \emph{suspicious} of its validity on operational grounds.

\subsubsection{Outline of the Derivation}
The reconstruction then proceeds as follows.  The goal is to obtain an expression for the overall transition amplitude~$\alpha$ from the two detections,~$\ell_1, \ell_2$, at~$t_1$ to the two detections,~$m_1, m_2$ at~$t_2$.  Suppose for the sake of argument that one adopts the view that there indeed exist two particles during the interval~$[t_1, t_2]$, labelled~$1$ and~$2$.  In that case, we know that one of two \emph{transitions} occurred in~$[t_1, t_2]$, unseen by us:~the \emph{direct} transition~(a particle---say, particle~$1$---travelled from~$\ell_1$ to~$m_1$, and particle~$2$ travelled from~$\ell_2$ to~$m_2$) or the \emph{indirect} transition~(particle~$1$ from~$\ell_1$ to~$m_2$, and particle~$2$ from~$\ell_2$ to~$m_1$).  These transitions can be assigned a Feynman transition amplitude,~$\alpha_{12}$ or~$\alpha_{21}$, respectively, which can be computed using standard methods.   However, since we have bracketed assertions about such transitions, we cannot directly combine the~$\alpha_{12}$ and~$\alpha_{21}$ via the Feynman sum rule.  This is, in fact, fortunate since application of the Feynman sum would yield~$\alpha = \alpha_{12} + \alpha_{21}$, which is \emph{not} empirically adequate~(it accounts for bosons but does not allow for the possibility of fermionic behaviour).

Instead, the reconstruction posits that, even though the assertion concerning unseen transitions is bracketed,~$\alpha$ is nevertheless determined by~$\alpha_{12}, \alpha_{21}$ via some function,~$H$, to be determined:
\begin{equation}
\alpha = H(\alpha_{12}, \alpha_{21}).
\end{equation}
To this posit is added an analogous assumption~(Eq.~(2) in~\cite[\S1]{Goyal2015}) for detections at three times~$t_1, t_2, t_3$.  Together, these comprise the \emph{operational indistinguishability postulate}~(OIP)\index{quantum reconstruction!identical particles!operational indistinguishability postulate}.

A noteworthy constraint on~$H$ is obtained by inspection of standard modelling heuristics.  Specifically, in a case where one has two identical particles in subsystems that are regarded as isolated from one another, such as two electrons in separate hydrogen atoms, one finds through application and testing of the quantum formalism that the fact of identicality is unimportant.  This leads to the \emph{isolation condition}\index{quantum reconstruction!identical particles!isolation condition}, namely that:
\begin{equation}
\left| H(\alpha_{12}, 0) \right| = \left| \alpha_{12} \right|
\end{equation}
and
\begin{equation}
\left| H(0, \alpha_{21}) \right| = \left| \alpha_{21} \right|.
\end{equation}

With the OIP and isolation condition in hand, various combinatorial symmetries in concert with the Feynman formalism for single systems yield the correct form of~$H$:
\begin{equation}
\alpha = \alpha_{12} \pm \alpha_{21},
\end{equation}
which generalizes to any number of particles:
\begin{equation}
\alpha = \sum_{\pi \in S_N} \left(\sgn{\pi} \right)^\sigma \alpha_{\pi}.
\end{equation}
Here, $S_N$ is the symmetric group of order~$N$,  and~$\alpha_{\pi}$ is the transition amplitude in the case where particles~$1, \dots N$ go from~$\ell_1, \dots, \ell_N$ to~$m_{\pi(1)}, \dots, m_{\pi(N)}$.   The free parameter~$\sigma$ can take the value~$0$ or~$1$, corresponding to bosons or fermions.

\paragraph*{State form of the Symmetrization Postulate.}
From the thus derived Feynman form of the symmetrization postulate, the state form is obtained~\cite[\S2]{Goyal2015}.  In the simplest case of two particles in one dimension,
\begin{equation}
\psi_{\textsc{\textsf{ID}}}(x_1, x_2) = \psi(x_1, x_2) \pm \psi(x_2, x_1) \quad\quad x_1 \leq x_2,
\end{equation}
where~$\psi(x_1, x_2)$ corresponds to the state of the system if one grants there exist two particles, whilst~$\psi_{\textsc{\textsf{ID}}}(x_1, x_2)$ is the actual state of the system.  This derivation throws the usual interpretation of the symbolism of the symmetrization postulate into question.  In particular, the physical referent of the indices~$i$ differ on the two sides of the equation. On the left-hand side, in the function~$\psi_{\textsc{\textsf{ID}}}(x_1, x_2)$, the referent of $x_i$ is not particle~$i$ but the~$i$th detection.  In the function~$\psi(x_1, x_2)$ on the right-hand side, $x_i$ refers to particle~$i$.  However, as the assumption of unseen transitions has been bracketed, the latter referent cannot be taken literally.  In short, the almost universal interpretational assumption that~$i$ is a \emph{particle label} is brought into question\index{quantum reconstruction!identical particles!particle-label interpretation of indices}.

\subsection{Interpretation} \index{interpretation!identical particles}\index{identical particles!interpretation!interpretation of reconstruction}

\subsubsection{Step 1:~Complementarity Object-Models}\index{identical particles!complementarity of persistence and non-persistence}
\label{sec:interpretation-step-1}

The above-summarized reconstruction \emph{brackets} assertions concerning unseen transitions.  The reconstruction demonstrates that, despite this bracketing, it is possible to derive the empirically-validated symmetrization postulate from the operational indistinguishability postulate~(OIP).  As mentioned in Sec.~\ref{sec:exploiting-a-gap}, the rationale given for this bracketing is that identical particles are `indistinguishable'.  However, if we take this to mean that identical particles are \emph{not reidentifiable}, this rationale implicitly presumes that the particles in question are persistent individuals~(since the (non-)reidentifiability of an object presumes its persistence).  But, if the identical particles are, in fact, persistent individuals at all times, why cannot one simply apply the Feynman sum rule to combine the~$\alpha_{\pi}$~(which, as mentioned earlier, excludes fermionic behaviour, and hence is not empirically adequate)?  Conversely, if the identical particles are absolutely \emph{not} persistent individuals, why does the OIP---which directly utilizes amplitudes~$\alpha_{\pi}$ computed on the assumption that the data \emph{is} generated by persistent individuals---yield an empirically-valid outcome?  The challenge of resolving this conundrum drives the first interpretative step.

The complementarity interpretation\index{interpretation!identical particles!complementarity of persistence and nonpersistence} reported in~\cite{Goyal2019a} asserts that the OIP should be understood as a mathematical \emph{synthesis} of two \emph{different object-models} of the same data.  That is, given detections~$\ell_1, \ell_2$ at~$t_1$ followed by detections~$m_1, m_2$ at~$t_2$, one can construct two \emph{different} object-models of this data.  In the first model, the \emph{persistence model}, one posits that there exist two persistent individuals that are responsible.  However, in the second model, the \emph{nonpersistence model}, one posits that there is in fact a \emph{single} individual which is responsible for the data.  This single individual could be regarded as a kind of holistic object~(whose nature is to be determined), which yields \emph{two} spatially-separated detections at each time.  

The essential idea here is that although we are wont to interpret the detection of two spatially separated events as due to two separate individuals~(as this accords with everyday experience of the physical world), one could \emph{logically} regard these events as the manifestation of a single object.  This echoes the tension between the metaphysical positions of pluralism and monism.  And in the microscopic context, where we do not directly observe the objects in question~(we only observe detections events), such a logical possibility ought not be casually dismissed.

As described in~\cite[\S4]{Goyal2019a}, these two models are naturally accommodated under the umbrella of \emph{complementarity}.  In short, neither model is unequivocally correct or incorrect.  Rather, they express partial truths, and only their synthesis yields an empirically adequate formalism.  Indeed, one can construct an illuminatingly close analogy to the diffraction of an electron at a double slit:~the Feynman sum rule that is employed in this context can be interpreted as a mathematical synthesis of a `wave' and `particle' model of an electron.

The upshot of the complementarity interpretation is that the very language of identical `particles' is misleading---a system of so-called identical `particles' is neither a collection of persistent individuals~(as the word `particles' suggests) nor a single persistent object.  Rather, \emph{both} of these ways of conceptualizing the underlying cause of the data is partially true, just as is the case for the so-called particle and wave models of the behaviour of a single electron.  Accordingly, their mathematical synthesis yields a formalism which cannot be interpreted as implying that the system \emph{is} either one or the other type of object~(\ie a composite of individuals, or a single individual).  This explains why one must bracket assertions about unseen transitions:~it is not a fact of the matter that there exist two persistent individuals in~$[t_1, t_2]$; hence one cannot presume that a direct or indirect transition actually occurred.

\subsubsection{Step 2:~Unconditional Individual Persistence: Loss of Empirical Cover} \index{interpretation!identical particles!individual persistence}
\label{sec:interpretation-step-2}

The second step is prompted by the following reflection on the complementarity interpretation.  Let us grant that the mathematical synthesis of two complementary object-models of the same data yield the symmetrization postulate.  This certainly provides \emph{post-hoc} justification for positing two such models.  However, is there a compelling \emph{a priori} reason for expecting that two such models would be necessary when modelling a system of  identical quantum particles?  In particular, why are two such models \emph{not} needed to model identical particles in \emph{classical} mechanics nor to model \emph{non-identical} particles in quantum mechanics?  In short, what is special about the confluence of identical particles and quantum mechanics that makes this object-model complementarity necessary?

\paragraph{Operational analysis of reidentification.}\index{interpretation!identical particles!operational analysis of reidentification} The answer to the above questions is obtained through an operational analysis of the process of reidentification of particles~\cite[\S2]{Goyal2023a}.  First consider point particles in classical mechanics\index{reidentification!in classical mechanics}.  In that context, a position measurement of a particle at an instant of time yields a point-valued outcome---a point in~$\numberfield{R}^3$.  The primary means for reidentifying a particle in the midst of others is to \emph{track} it---to perform position measurements of sufficiently high resolution and frequency.  Since such ideal position measurements are \emph{passive}~(\ie do not influence the state of the particle), an experimenter can probe the actual trajectory as precisely as desired\index{reidentification!by tracking}.   

However, a secondary means of identification---\emph{property-based reidentification}---is also available\index{reidentification!by time-independent properties}.  This relies on the fact that classical particles have time-independent properties, such as mass and charge.  If one measures such properties of a particle initially, then one can reidentify this particle later by simply measuring these properties, \emph{provided} that this particle is not identical~(in the values of these time-independent properties) to any other.  However, if one is dealing with a system of identical classical particles, only \emph{tracking-based reidentification} is available.

In the quantum context\index{reidentification!in quantum mechanics}, the key difference in the analysis arises from the fact that quantum measurements are \emph{active} processes.  This is taken into account in experimental design through \emph{segregating} measurements and (unitary) interactions.  For example, if an electron is in a magnetic field, one cannot design an experiment to track the electron through arbitrarily high-frequency, high-resolution position measurements, for such measurements would interfere with the electron's interactions with the magnetic field.  However, for a sufficiently isolated particle, coarse tracking enables reidentification.  Through such tracking, one can also measure the values of the time-independent properties of the particle.  In summary, one can reidentify~(with caveats) through coarse tracking and through initial- and final-measurements of time-independent properties.

However, in the case of a system of two identical quantum particles, property-reidentification is manifestly unavailable.  Is tracking-based reidentification available?  Well, if the particles are sufficiently separated, then tracking is possible through coarse position measurements.  But, if, say, two electrons enter into a collision, then there is a region of space in which performing sufficiently high-resolution, high-frequency position measurements would interfere with the collisional interaction.  Thus, tracking throughout the collisional process is an impossibility.  That is, one can track the two electrons on their way into, and way out of, the collisional region, but not within the region itself.  

Now, one ordinarily assumes that the two electrons continue to exist as persistent individuals within the collisional region.  But the above analysis shows that we no longer have any means to reidentify the electrons since we cannot track them through this region.  Therefore, the metaphysical assertion of individual persistence has lost empirical cover---the metaphysical assertion no longer has the backing of a performable experimental procedure, \viz. reidentification.   In contrast, if one is watching two aeroplanes, and these then disappear into cloud and then reappear a while later, the assertion that they continued to exist even whilst obscured from view is supported by the \emph{possibility} that one could have performed experiments~(\eg radar sounding) to confirm their continued individual existence during their visual obscuration.

Hence, returning to the set-up in the reconstruction~(\S\ref{sec:exploiting-a-gap}), where one observes detections~$\ell_1$ and~$\ell_2$ at~$t_1$ followed by~$m_1$ and~$m_2$ at~$t_2$:~the metaphysical assertion that there exist two persistent individuals in the intervening period \emph{in general} lacks empirical cover.  Clearly, there are instances where one \emph{can} justify this assumption, such as with two identical particles in separate laboratory experiment~(\eg where each is confined to a separate particle trap).  But we  have clear reason to question the \emph{general} validity of this metaphysical claim, while at the same time recognizing that there must be circumstances~(limiting cases) where this claim yields empirically-valid predictions.

Accordingly, the dual object-models can be regarded as a formal means of \emph{responding} to the bracketing of the metaphysical assumption of \emph{unconditional} individual persistence\index{persistence!individual persistence!bracketing in systems of identical particles} by allowing the validity of this assumption to depend upon the physical context.  This is achieved by one of the models---the nonpersistence model---positing the persistence of a \emph{single} object, rather than two.

On this basis, we can readily understand why a duality of object-models is \emph{not} required in systems of identical classical particles.  Since classical measurement is \emph{passive}, precise tracking is always possible, so that the assumption of individual persistence is always backed by the possibility of exact reidentification.  Similarly, in systems of non-identical quantum particles, the assumption of individual persistence is empirically well-supported since reidentification is always possible on the basis of pre- and post-interaction measurements of intrinsic properties.

\subsubsection{Step 3:~Identical Particles as Potential Parts}\index{interpretation!identical particles!identical particles as potential parts}\index{identical particles!identical particles as potential parts}

The operational analysis of Step~2 provides a clear operationally-grounded justification for questioning the validity of the metaphysical posit of unconditional individual persistence in the context of systems of identical quantum particles.  This provides a clear \emph{a priori} rationale for modelling the experimental data via complementary object-models.  However, the complementary object-models raises the question of how one can speak meaningfully in metaphysical terms about microscopic `objects' such as electrons.  In short, if one cannot unconditionally think of these as individually persistent entities, nor even as manifestations of a single holistic object, how can one usefully think and speak about them?  Is there a conceptual framing which does justice to the subtlety revealed by the complementary object-models?

Thus, in Step~3 of the interpretational process, we adopt a metaphysical stance, and seek appropriate metaphysical conceptual frameworks which illuminate what has been uncovered in the previous steps.  We first recognize that \emph{both} complementary object-models refer to the \emph{same} data.  That is, whereas one cannot say that electrons \emph{are} described by one or other object-model since neither of these models is by itself empirically adequate, the detection events are treated as absolute.  This is consistent with an event ontology, which accords primary reality to the event data~(in the sense that such data is \emph{classically definite}) and treats objects as secondary constructs for ordering event-data.

However, the empirical success of the symmetrization postulate, derived here through a mathematical synthesis of complementary object-models, supports the claim that there \emph{is} a reality that underpins these detection events.  The question then becomes how we can meaningfully speak of this underlying reality, even if not in the familiar, classical terms of unconditionally persistent individuals.  

Here, we appeal to the doctrine of \emph{potential parts}~\cite{Holden2006}, which addresses the following fundamental question: when an object is divided into two parts, did those two parts exist prior to the division?  The answer hinges on how one conceives of the process of division.  Is it a \emph{passive} process, which merely \emph{separates} two parts that were previously present?  Or is it an \emph{active} process, which \emph{co-creates} the parts from the original whole?  The doctrine of potential parts asserts that division is an active process, and accordingly holds that the parts did \emph{not} exist prior to the cutting.  In contrast, the doctrine of actual parts asserts the contrary, \viz division is mere separation, and the parts exist even prior to division.  The doctrine of actual parts is the metaphysical basis for the doctrine of atomism, and is woven into classical physics.  However, here we have reason to believe that the doctrine of potential parts is applicable.

In light of the doctrine of potential parts, we can speak of a system of identical particles as consisting of identical particles \emph{potentially}, not \emph{actually}.  Here, `\emph{potentially}' expresses the idea that there are contexts in which two identical particles \emph{as individuals} will emerge.  And `not \emph{actually}' expresses the idea that, \emph{in general}, one cannot view the system as consisting as an aggregate of individuals~(as one would conceive classically, given the atomistic conception of matter).  Thus, the two electrons in a helium atom exist potentially, not actually---after complete ionization, there would exist two electrons actually~(in the sense that the isolation condition holds, and the persistence model by itself is empirically adequate).

%
%

\subsection{Limitations of the Prevailing Methodology} \index{interpretation!identical particles!limitations of prevailing methodology}\index{identical particles!interpretation!limitations of prevailing methodology}

As described in the Introduction, the prevailing methodology for interpreting quantum theory begins with the quantum formalism itself.  The same is true is the particular case of systems of identical particles.  As mentioned in the Introduction, such an approach is inherently problematic.  The reconstruction-based interpretation of identical particle systems, which is summarized above, helps to bring into focus the types of pitfalls that one faces.  These pitfalls are analysed in detail elsewhere~\cite[\S2]{Goyal2025b}, so I will confine myself to a few brief remarks.

\subsubsection{What is the `Identical Particle Quantum Formalism'?}
First, although there is no dispute~(at least as far as I am aware) about what the abstract quantum formalism consists of, there is very little consensus on precisely what are the quantum rules for handling identical particles.  There is broad agreement that the symmetrization postulate is part of these rules.  But reflection on the process of constructing a quantum model of a helium atom, or a box of helium gas, makes clear that a number of other steps are involved which invoke~(implicitly or explicitly) nontrivial physical ideas or assumptions.

For example, one starts the modelling process by constructing a \emph{classical} model of the atom, in particular writing down a \emph{classical} Hamiltonian~$H$.  Then one constructs a quantum \emph{foil} model by `quantizing' the classical Hamiltonian to obtain a quantum hamiltonian,~$\hat{\mathsf{H}}$, and solving for the energy eigenstates.  However, this model is not yet empirically valid since no special account has yet been taken of the identicality of electrons.  To rectify this, the symmetrization postulate is invoked, which yields empirically-valid anti-symmetric wavefunctions, and measurement operators are restricted to those that are symmetric.  In addition, implicitly, such a modelling process invokes an isolation condition since it chooses to ignore all other electrons in the universe, such as those belonging to other helium atoms in the box of gas.  In short, the identical particle formalism is best viewed as an \emph{algorithm}---a multi-step procedure, of which the application of the symmetrization postulate is only one step.    However, as far as I am aware, prior to~\cite[\S2]{Goyal2025b}, this procedure has never been explicitly laid out.  In textbooks, many steps are left implicit, being \emph{shown} by example rather than clearly articulated.

In the interpretation of the state-based identical particle formalism, philosophical reflection is usually directed to the symmetrization postulate alone.  However, there is no \emph{a priori} reason to suppose that the other parts of the algorithm are not interpretatively relevant.  Indeed, only the entire algorithm has empirically warrant.   The neglect of the other parts of the algorithm seems to reflect the general tendency to marginalize modelling heuristics and to focus on formalism that appears crisp and general.

The reconstruction-based interpretation side-steps these issues.  The reconstruction effectively distills the content of the identical particle formalism into the OIP and isolation condition.  Hence, if one focusses interpretational attention on these~(in the context of Feynman's rules for single systems), one can be assured of having taken into account most of what is interpretationally relevant.  As we have seen above, what remains can be subsequently unveiled by posing appropriate questions and performing operational analysis, logical analysis of concepts, and metaphysical reflection.

\subsubsection{Minimal Operational Interpretation of the Symbolism}

The mathematical formalism of a theory is invariably accompanied by a minimal physical interpretation of the symbolism.  This cannot be avoided:~the minimal physical interpretation is invoked when the formalism is used to build models of physical systems.  In the case of the identical particle formalism, the indices~$i$ in symmetrized states such as~$\psi(x_1, x_2) = \left[ \alpha(x_1)\beta(x_2) \pm \alpha(x_2)\beta(x_1)\right]/
\sqrt{2}$ are minimally interpreted as \emph{particle} labels.  However, careful analysis shows that this interpretation is not actually needed at all steps of the above-mentioned algorithm.  Rather, this interpretation is only necessary up to~(and including) the construction of the above-mentioned foil model.

However, until recently, the particle-label interpretation of indices has been near-universally assumed in interpretational work.  Only in recent years~(\eg \cite{DieksLubberdink2020,CaultonPhDthesis}) has this assumption been seriously questioned.  However, such questioning has yet to yield a well-founded \emph{alternative} interpretation of the indices.  

Again, the reconstruction-based interpretation avoids such ambiguities.  Indeed, the reconstruction not only shows that the above-mentioned authors were correct in questioning the standard interpretation of the indices, but the complementarity interpretation of the reconstruction provides an explicit alternative interpretation of them.

\section{Concluding Remarks}

The creation of quantum theory was a watershed moment in the noetic evolution of our species.  However, its potential for transforming our patterns of thought and our understanding of nature has yet to be fully realized.  A key step in its fulfillment is the development of a fully worked out quantum conception of reality.  A conception that enjoys widespread assent is important not only in physics and philosophy, but far beyond the academy.  

In this paper, I have argued that the prevailing interpretative methodology is ill-suited to the task of interpreting quantum theory, for two main reasons.  First, this methodology neglects the theoretic content implicit in modelling heuristics and experimental practices, as well as the theoretic content encapsulated in the mathematical structures of the quantum formalism.  Second, it offers little protection against undue influence from metaphysically-laden language that invariably accompanies the formalism.

As a remedy, I have proposed an alternative methodology, which aims to take into account all relevant theoretic content.  In particular, this methodology involves the construction of an interpretation-free zone around the theory, wherein one adopts a descriptive stance, with the goal of surveying all theoretic content without bias.  The methodology harnesses the results of the quantum reconstruction program to render the content of the quantum formalism philosophically digestible.  It also makes extensive use of operational analysis to trace back concepts as far as possible to the `bare data' of laboratory experience, with the goal of uncovering unwarranted extrapolation of deeply engrained concepts into our understanding of the microworld and of discovering or formulating concepts that are better fitting.

As a case study, I have summarized the reconstruction of the identical particle formalism and its step-by-step interpretation.  This has lead to \begin{inlinelist} \item a new principle of complementarity applicable to systems of identical particles; \item a clear operationally-grounded understanding for the need to bracket the assumption of individual persistence in systems of identical particles; and \item the novel metaphysical proposal that identical particles be considered potential parts of a whole. \end{inlinelist}

\begin{acknowledgements}
I am grateful to the editors of this volume, Lars-G\"oran Johansson and Jan Faye, for their illuminating correspondence on the some of the key topics discussed herein.  I also thank Lars-G\"oran for helpful remarks on an earlier version of this article.
\end{acknowledgements}

\newpage

\bibliographystyle{plainnat}

\bibliography{references_master}

\printindex

\end{document}